# FACEBOOK IMPLEMENTATION IN DEVELOPING ENGLISH WRITING FOR THAI STUDENTS

Pisutpong Endoo [1]

[1] Linguistics Department, Faculty of Management Technology, Rajamangala University of Technology Isan, Surin Campus, Thailand
http://www.surin.rmuti.ac.th/surin
promise_guy@hotmail.com

*ABSTRACT*

*The objectives of this research were to study FB implementation and attitudes in developing English writing skills of Thai students studying in the first year students program in EIC academic year 1/2014 at RMUTI, Surin Campus. The Purposive sampling was designed for data collecting. The instruments for this research were questionnaires and in-depth questions. The data analysis was analyzed by the Descriptive statistics to find out the value of the frequency and percentage.*

## 1. INTRODUCTION

Thailand is formerly known as Siam which is in Southeast Asia. It is bordered to the north by Burma and Laos, to the east by Laos and Cambodia, to the south by the Gulf of Thailand and Malaysia, and to the west by the Andaman Sea and the southern extremity of Burma. Its maritime boundaries include Vietnam in the Gulf of Thailand to the southeast, and Indonesia and India on the Andaman Sea to the southwest [1]. Thailand has a total area of approximately 513,000 km2 (198,000 sq mi) and around 66 million people. Bangkok is the capital city. The country's official language is Thai and the primary religion is Buddhism, which is practiced by around 95% of the population. Among the ten ASEAN countries, Thailand ranks second in quality of life [2].

Education in Thailand in a brief is provided mainly by the Thai government through the Ministry of Education from pre-school to senior high school. A free basic education of twelve years is guaranteed by the constitution, and a minimum of nine years' school attendance is mandatory. Formal education consists of at least twelve years of basic education, and higher education. Basic education is divided into six years of elementary education and six years of secondary education, the latter being further divided into three years of lower- and upper-secondary levels. Kindergarten levels of pre-elementary education, also part of the basic education level, span 2–3 years depending on the locale, and are variably provided. Non-formal education is also supported by the state. Independent schools contribute significantly to the general education infrastructure. Administration and control of public and private universities are carried out by the Office of Higher Education Commission, a department of the Ministry of Education [3].

Studying English in Thailand had systematically started since 1836. That time English class was educated at the Buddhist temple, Wat Bowonniwet Vihara, in Bangkok now. After that it has been made a class in the both public and government of schools, collages, and universities around Thailand [4].

Although English class has been studied and taught for Thai students for a long time, Thai students' effective English still have many problems. They cannot confidentially communicate with other people can speak, write, read and listen to English [5]. Both Thai students and teachers should find out the new technical ways to learn and teach English especially always changing modern world [5].

At the present, internet is very influential and important for the people until it becomes their part of life. It is used for education, working, communication, entertainment, and recreation etc. Internet is telecommunication system connecting with all computers in order to communicate between internet users. It can be said that now it is the widest network [6]. The communication is connection between messengers consisting of remained context or information with message receivers via online social network system. Messengers and message receivers belong to this network. Now this system has continuously grown in the Thai society and international countries; for example, FB, Twitter, YouTube, Google Plus, Instagram etc. All of these have become new media and source used for information searching [7].

With the modern technology innovated very quickly and lifestyle of the city people, it has changed very much. Most of people use technology all the time. They can very easily used internet to share new update knowledge and to communicate with freedom through the social network. FB is a part of those networks used for communication. It is very popular network because it can be used with the both synchronous activities and none synchronous activities. FB can be also used for supporting knowledge activities for teachers [8]. Social network had originated since 1969 and become very popular for American teenagers since 1997. Later members of users have continuously increased until now there are 85 % of the population around the world can access the social network system especially FB [9]. FB was found by Mark Zuckerburg with his close friends, Dustin Moskowitz and Christ Hughes on February 4, 2004. It was released for general users in 2006 and in 2010 there were users about 400 billion per a month and become the biggest online social network in the world now [10].

According to the survey of FB usage in Thailand, it was found that there were Thai people used it 16.1 million in 2008 [7]. A group very much used FB was the both male and female teenagers during 18-25 years old, and the later the both male and female adults during 26-34 years old respectively. However, the teenagers during 13-17 years old were the most skillful in digital technologies. Now FB has becomes very necessary network and main factor in the university. The students, teachers and other staffs have own account users and they access it to search information and use it for class teaching implementation [11]. FB network system is very easy to use and has varieties of functions implemented for working and teaching in the university. With this reason, it becomes very impact for staffs in the university [12].

Regarding to the information above, so the researcher was interested to study FB Implementation in Developing English Writing Skills: A Case Study of First Year Students Program in EIC academic Year 1/2014 at RMUTI, Surin Campus. The objectives of this study aimed to study FB implementation and attitudes in developing English writing skills of the first year students program in EIC academic year 1/2014 at RMUTI, Surin Campus.

## 2. MEDOLOGY

This research was class research. The objectives of this were research to study FB implementation and attitudes in developing English writing skills of the first year students program in EIC academic year 1/2014 at RMUTI, Surin Campus. The Purposive sampling was designed for data collecting. There were 53 students studying of the first year program in EIC academic year 1/2014 at RMUTI, Surin Campus. The instruments for this research were questionnaires. The questionnaire was divided into three parts consisted of part 1: The General demographics of population, Part 2: FB implementation in developing English writing skills of population and Part 3: Attitudes on FB implementation in developing English writing skills and recommendation of population. The data analysis was analyzed by the Descriptive statistics to find out the value of the frequency and percentage.

## 3. RESULTS & DISCUSSION

### 3.1. Results of the Study

From the study, the findings were found as these followings:

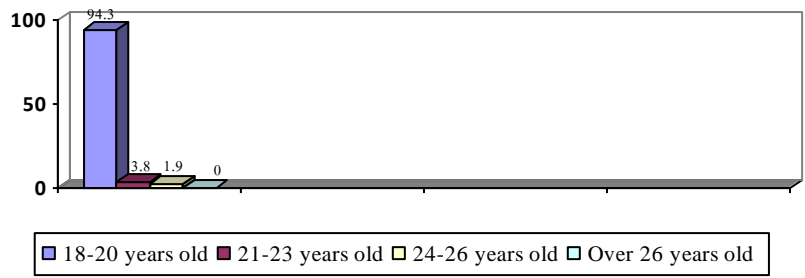

Figure 1: General demographics of population's age

The bar graph 1 showed the general demographics of population's age. There were 94.3 % of populations during 18-20 years old, 3.8 % of population during 21-23 years old and 1.9 % of population during 21-23 years old. There were not students during over 26 years old.

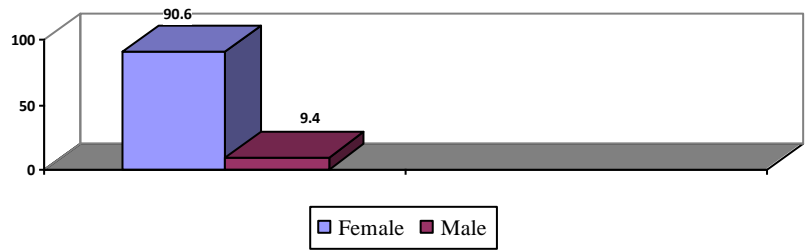

Figure 2: General demographics of population's sex.

The bar graph 2 showed the general demographics of population's sex. 90.6 % of populations were female and 9.4 of populations were male respectively.

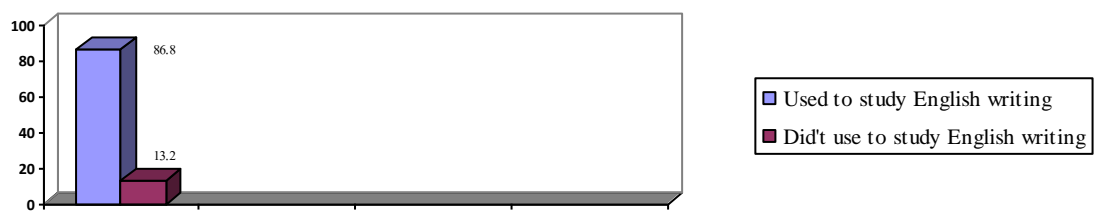

Figure 3: Experiences studying English writing

The bar graph 3 showed result of FB implementation in developing English writing skills of population about the experiences studying English writing at former school before they become the students in the university. According to this bar graph, there were 86.6 % of populations used to study English writing and 13.2 % of populations did not study English writing before respectively.

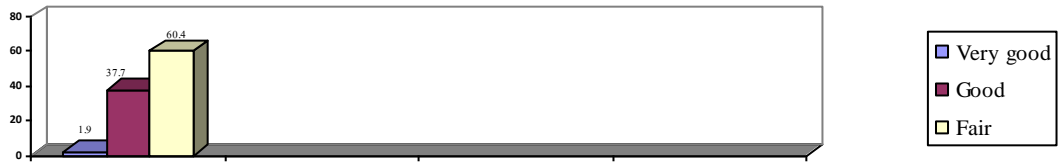

Figure 4: Populations' English writing capability

The bar graph 4 showed result of FB implementation in developing English writing skills of population about the populations' English writing capability. According to this bar graph, 60.4 % was fair, 37.7 % was good and 1.9 % was very good in English writing capability respectively.

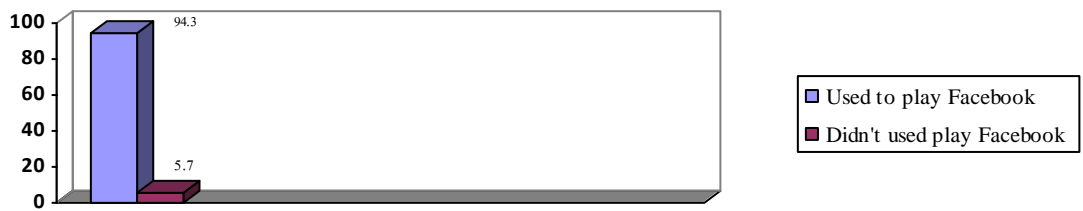

Figure 5: Experiences of playing FB

The bar graph 5 showed result of FB implementation in developing English writing skills of population about the experiences of playing FB at former school before they become the students in the university. According to this bar graph, there were 94.3 % of populations used to play FB and 5.7 % of populations did not use FB before respectively.

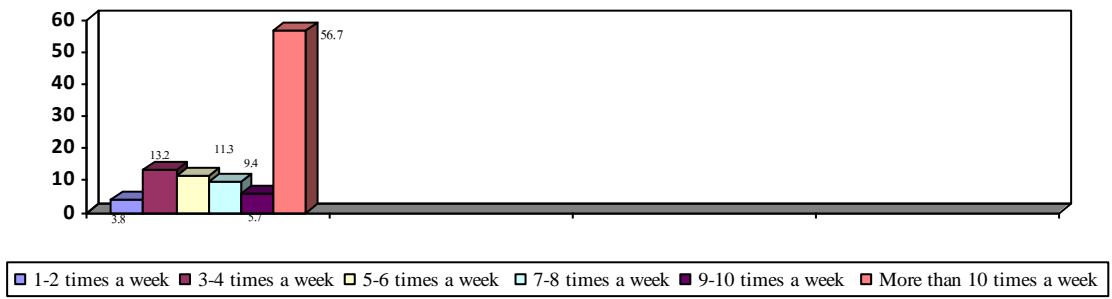

Figure 6: The frequency of FB playing

The bar graph 6 showed result of FB implementation in developing English writing skills of population about the frequency of FB playing per a week. According to this bar graph, there were three highest level identified the frequency of FB playing per a week. There were 56.7 % of populations identified that they played FB more than 10 times a week, 13.2 % played 3-4 times a week, 11.3 % played 5-6 times a week used to play FB and 5.7 % of populations did not use FB before respectively.

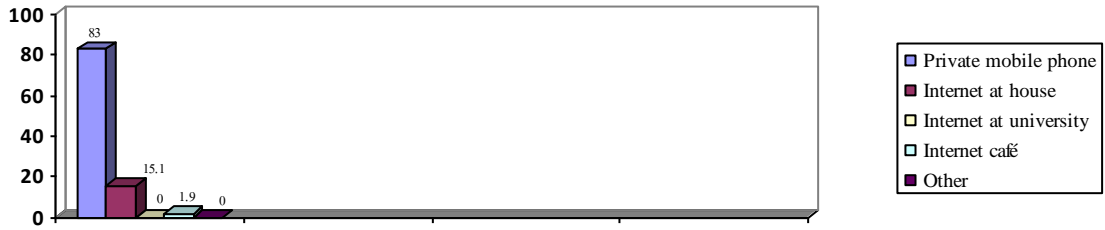

Figure 7: The place of population's FB playing

The bar graph 7 showed result of FB implementation in developing English writing skills of population about the place of population's FB playing. According to this bar graph, there were three highest level identified the place of FB playing. There were 83.7 % of populations identified that they played FB through their private mobile phone, 15.1 % played internet at their houses and 1.9 % played at internet café.

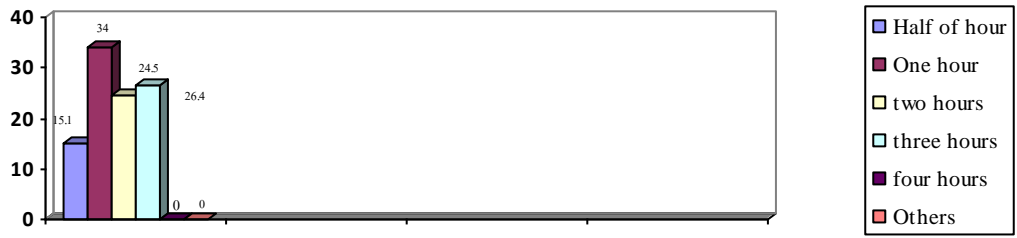

Figure 8: Number of hours of populations' playing FB

The bar graph 8 showed result of FB implementation in developing English writing skills of population about number of hours the population played FB. According to this bar graph, there were three highest level identified the number of playing FB of populations. There were 34.7 % of populations identified that they played FB one hour a time, 26.4 % played FB 3 hours a time and 24.5 % played FB 2 hours a time respectively.

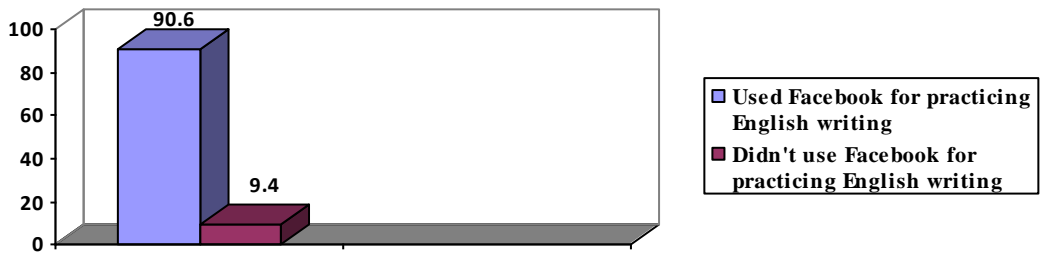

Figure 9: Using Fecebook for practicing English writing

The bar graph 9 showed result of FB implementation in developing English writing skills of population about using Fecebook for practicing English writing. According to this bar graph, there were 90.6 % of populations identified that they used FB for practicing English writing and 9.4 % didn't use it for practicing English writing respectively.

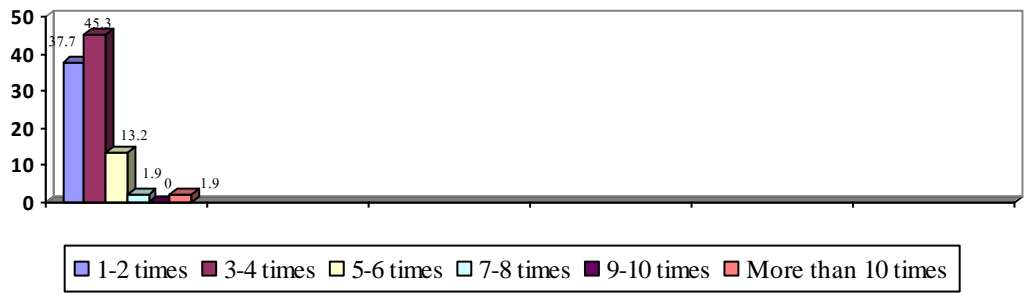

Figure 10: The number of time using FB implementation in developing English writing skills per a week

The bar graph 10 showed result of FB implementation in developing English writing skills of population about the number of time using FB implementation in developing English writing skills per a week. According to this bar graph, there were three highest level identified the number of time using FB implementation in developing English writing skills of populations per a week. There were 45.3 % of populations identified that they played FB 3-4 times a week, 37.7 % played if 1-2 times a week and 13.2 % played it 5-6 times a week respectively.

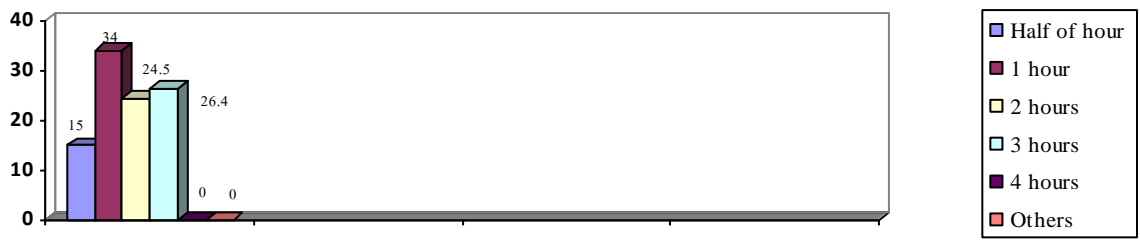

Figure 11: The number of hours using FB implementation in developing English writing skills per a week.

The bar graph 11 showed result of FB implementation in developing English writing skills of population about the number of hours using FB implementation in developing English writing skills per a week. According to this bar graph, there were three highest level identified the number of hours using FB implementation in developing English writing skills of populations per a week. There were 34 % of populations identified that they played FB 1 hour a week, 26.4% played it 3 hours a week and 24.5 % played 2 it per a week respectively.

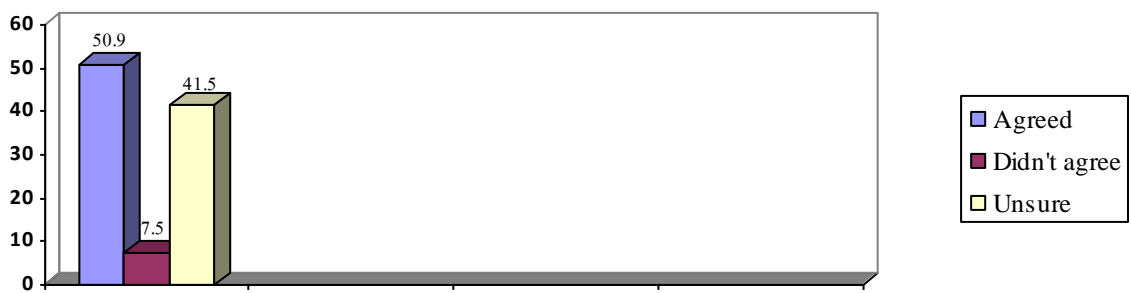

Figure 12: FB implementation in developing populations' English writing skills

The bar graph 12 showed result of FB implementation in developing English writing skills of population can develop English writing skills. According to this bar graph, there were 50.9 % of populations agreed with using FB implementation can develop English writing skills, 41.5 % didn't agree with this and 7.5 % was unsure respectively.

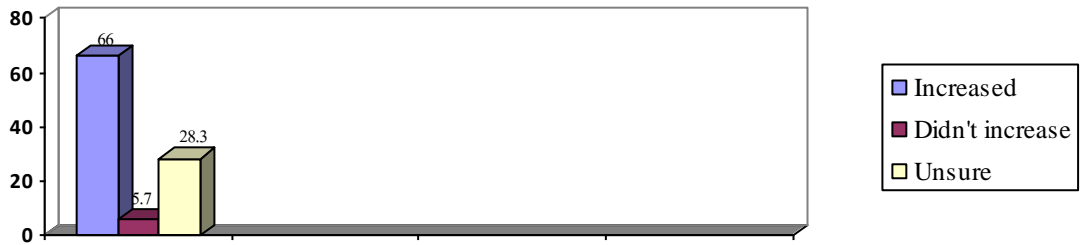

Figure 13: Increasing capability of English punctuation skill

The bar graph 13 showed result of FB implementation in developing English writing skills of population can increase the capability of English punctuation skill. According to this bar graph, there were 66 % of populations increased, 28.3% was unsure and 5.7 didn't increase the capability of English punctuation skill respectively.

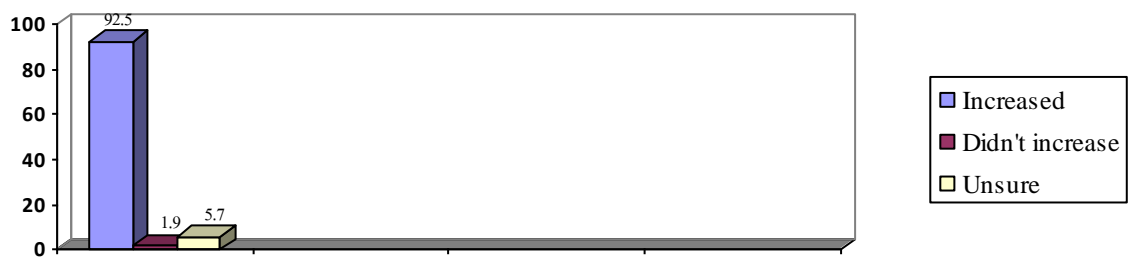

Figure 14: Increasing new English vocabularies

The bar graph 14 showed result of FB implementation in developing English writing skills of population can increase new English vocabularies. According to this bar graph, there were 92.5 % of populations increased, 5.7% was unsure and 1.9 % didn't increase the capability of new English vocabularies respectively.

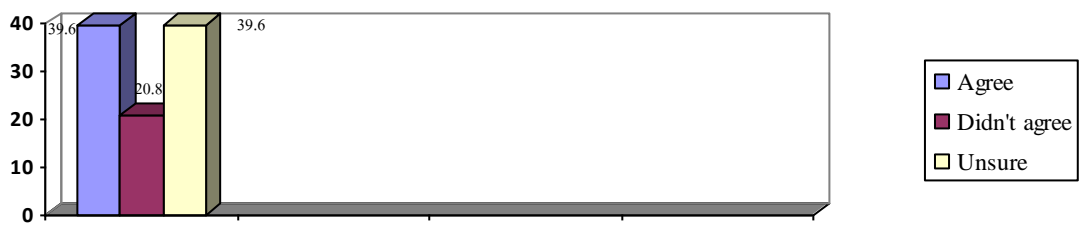

Figure 15: Practice to use the English structures of Tense

The bar graph 15 showed result of FB implementation in developing English writing skills of population can practice using the English structures of Tense. According to this bar graph, there were both 39.6 % agree and disagree with this point and 20.8 % was unsure respectively.

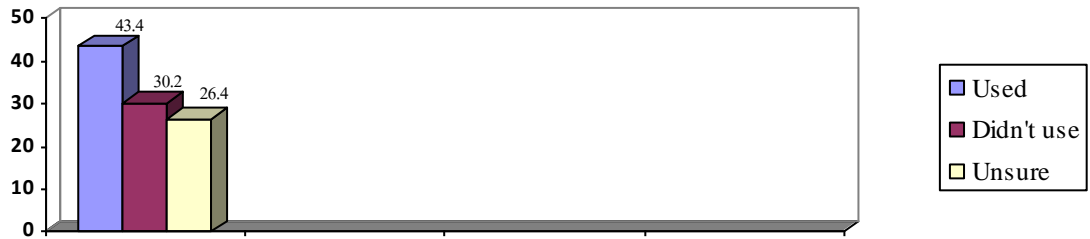

Figure 16: Spelling Check while FB implementation in developing English writing skills

The bar graph 16 showed result of using Spelling Check while FB implementation in developing English writing skills of population. According to this bar graph, there were 43.4 % used it, 30.2 % didn't used it and 26.4 % was unsure respectively.

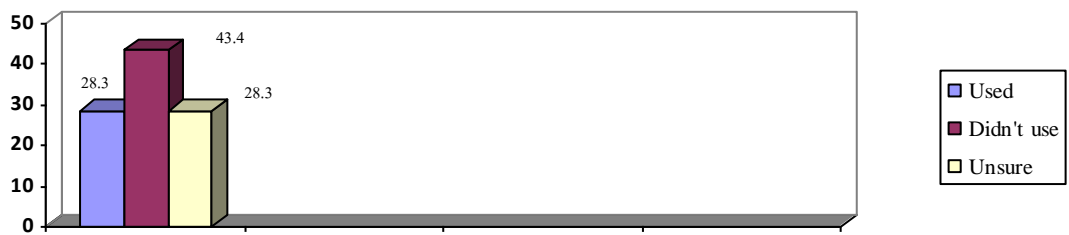

Figure 17: Using Grammar Check

The bar graph 17 showed result of using Grammar Check while FB implementation in developing English writing skills of population. According to this bar graph, there were 43.4 % didn't use it, 28.3 % used it and 28.3 % was unsure respectively.

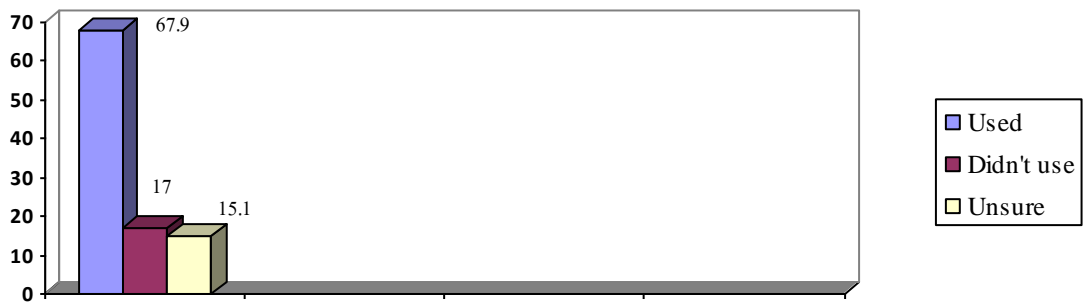

Figure 18: using Dictionary Check

The bar graph 18 showed result of using Dictionary Check while FB implementation in developing English writing skills of population. According to this bar graph, there were 67.9 % used it, 17 % didn't use it and 15.1 % was unsure respectively.

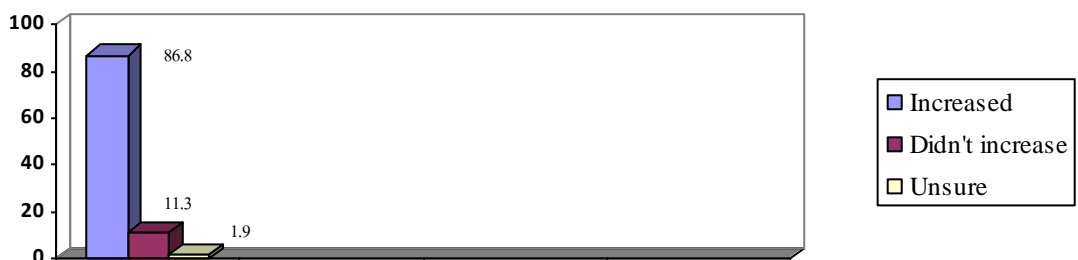

Figure19: Using new modern technology skill

The bar graph 19 showed result of FB implementation in developing English writing skills of population can increase using new modern technology skill. According to this bar graph, there were 86.8 % of populations increased, 11.3% didn't increase and 1.9 % was sure respectively.

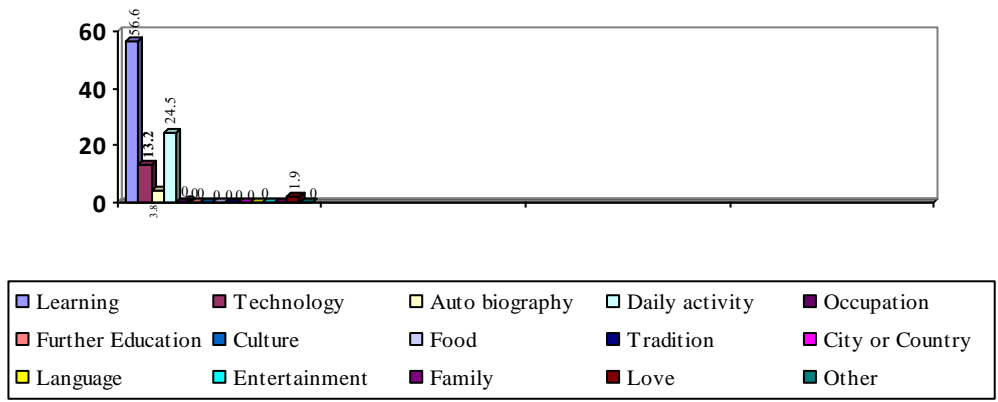

Figure 20: Used topics for FB implementation in developing English writing skills

The bar graph 20 showed result of type of topics the populations used FB implementation in developing English writing skills per a time. According to this bar graph, there were three highest topics the populations used FB implementation in developing English writing skills of populations. There were 56.6 % of populations identified that they liked to write about Learning, 24.5 % about Daily Activity and 13.2 % about Technology respectively.

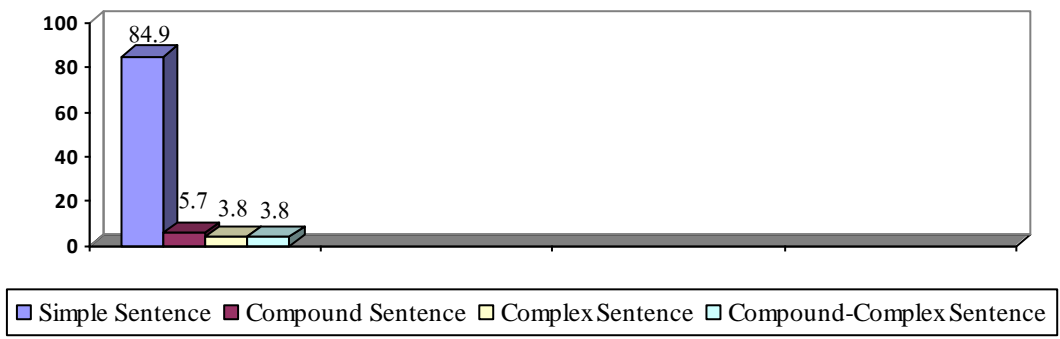

Figure 21: Types of sentences which were very often used

The bar graph 21 showed result of sentences types the populations very often used FB implementation in developing English writing skills of population per a time. According to this bar graph, there were three highest sentence types the populations very often used FB implementation in developing English writing skills. There were 84.9 % of populations very often used Simple sentence, 5.6 % used Compound Sentence and 3.8 % used Complex Sentence and Compound-Complex Sentence respectively.

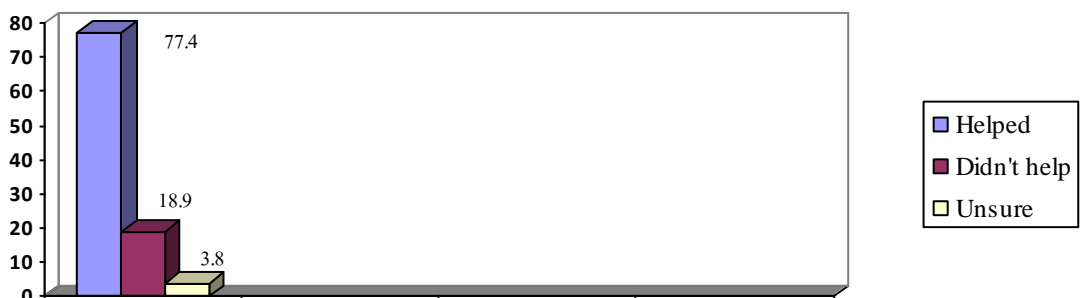

Figure 22: FB can develop English writing skills

The bar graph 22 showed result of attitudes on practicing English writing through FB can develop English writing skills. According to this bar graph, there were 77.4 % of populations could help, 18.9 % couldn't help and 3.8 % was unsure that practicing English writing through FB could develop English writing skills.

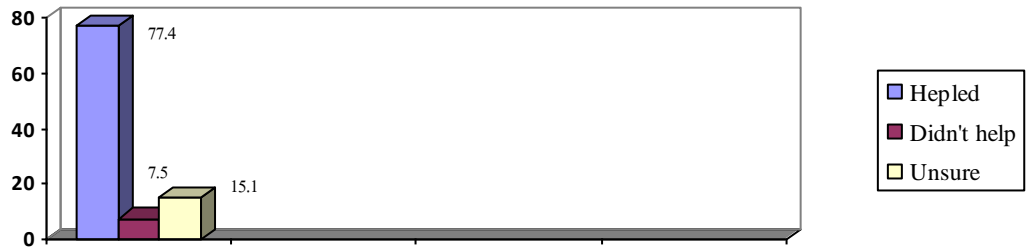

Figure 23: FB helped to like better English writing

The bar graph 23 showed result of attitudes on practicing English writing through FB helped to like better English writing. According to this bar graph, there were 77.4 % of populations helped, 7.5 % couldn't help and 15.1 % was unsure that practicing English writing through FB could help them like English writing.

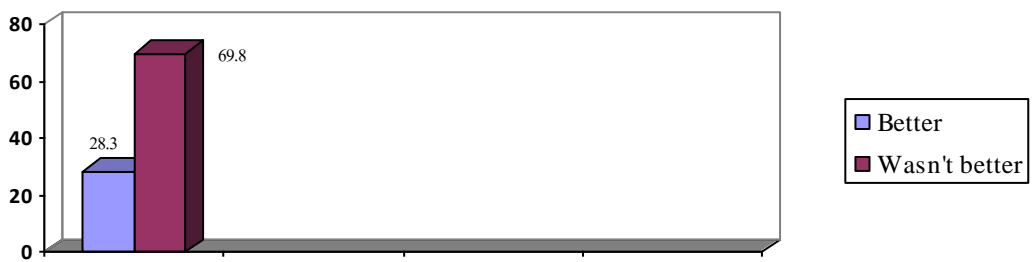

Figure 24: Practicing English writing through FB was better than practicing English writing with handwriting

The bar graph 25 showed result of attitudes on practicing English writing through FB was better than practicing English writing with handwriting. According to this bar graph, there were 69.8% of populations felt that it wasn't better and 28.3 % felt that practicing English writing through FB wasn't better than practicing English writing with handwriting

## 4.2 Discussion of the Result Study

According to results of the study, there was some point not relative with the research found by some scholar such as Supasapon's research [5]. She studied Secondary School Students' Internet Consuming Behavior in Daily Life: A Study of Traim-Udom Patanakarn School and found that most students used internet at their home spending 2 hours a time. However, this research was found that most populations used for this study used smart phones or their private mobile phones and spent an hour per time. There were 34 % of all populations used for this study.

Another very important point, most populations used for this study didn't agree with practicing English writing through FB network can be better skill than practicing it with their hand writing. From this result, it made us know that most students like to practice English writing with handwriting better than practicing writing through Fecebook network. There were 69.8 of all populations identified and supported this idea.

Lastly 92.5 % of populations identified that they were very confident and believable that FB implementation in developing English writing could increase new English vocabularies and they could receive a lot of knowledge in English. By the way this FB implementation will develop their English writing skill in the future.

## 5. CONCLUSION

The objectives of this research to study Facebook implementation and attitudes in developing English writing skills of the first year students program in English for International Communication academic year 1/2014 at Rajamagala University of Technology Isan, Surin Campus.

The Purposive sampling was designed for data collecting. There were 53 students studying of the first year program in English for International Communication academic year 1/2014 at Rajamagala University of Technology Isan, Surin Campus. The instruments for this research were questionnaires. The data analysis was analyzed by the Descriptive statistics to find out the value of the frequency and percentage.

The findings were found these followings:

1) The overall result of Facebook implementation in developing English writing skills of the first year students program in English for International Communication academic year 1/2014 at Rajamagala University of Technology Isan, Surin Campus was found that the background of Facebook implementation before being the students in the university was at the highest level (94.3 %), the later was Facebook implementation in developing English writing made the students learn new English vocabularies (92.5%) and Facebook implementation in developing English writing made the students improve English writing skill (90.6 %) respectively.

2) The overall result of attitudes on Facebook implementation in developing English writing skills of the first year students program in English for International Communication academic year 1/2014 at Rajamagala University of Technology Isan, Surin Campus was found that the students agreed that the Facebook implementation in developing English writing skills could improve their English writing skill, the later was Facebook implementation in developing English writing skills was better than practical writing through books (69.8) and were at the highest level (77.4), the Facebook implementation in developing English writing skills could integrate tense study in the future (37.7 %) respectively.

## 6. RECOMMENDATION

In a brief suggestion, although this research is completed, there are many interesting points to do the future research. And this research is only the first step to receive the basic data about the population. This will lead to create next lessons for them to practice English though Feacbook.

**ACKNOWLEDGEMENTS**

This work was supported by Rajamangala University of Technology Isan (RMUTI) funding academic year 2014. I would like to say a lot of thanks for this university's financial support.

## AUTHOR

Pisutpong Endoo is a head and lecturer of Linguistics Department, Faculty of Management Technology, Rajamangala University of Technology Isan, Surin campus, Thailand. He received his B.A. in English from Mahachulalongkornrajavidyalaya University, Chiang Mai campus in 2006, M.A. in 2008 and PhD. in 2012 in Linguistics respectively from University of Mysore, Karnataka, India. His areas of interest include English teaching, Linguistics, sociolinguistics, folk linguistics, anthropological linguistics in Thailand, and Thai cultural terms etc.

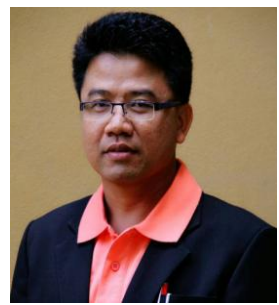